\documentclass[aps,physrev,twocolumn,10pt,superscriptaddress]{revtex4-2}

\usepackage[utf8]{inputenc}
\usepackage[english]{babel}
\usepackage[T1]{fontenc}
\usepackage{color}
\usepackage{xcolor}
\usepackage{microtype}
\usepackage{enumitem}
\usepackage[normalem]{ulem}
\usepackage[pdftex]{graphicx}
\usepackage{hyperref}
\usepackage{amsmath}
\usepackage{amssymb}
\usepackage{mathtools}
\usepackage{mathrsfs}
\usepackage{multirow}
\usepackage{bbm}
\usepackage{bbold}
\usepackage{braket}
\usepackage{siunitx}
\usepackage[only,llbracket,rrbracket]{stmaryrd}
\usepackage[caption=false]{subfig}
\usepackage{float}
\usepackage{makecell}
\usepackage[capitalize]{cleveref}

\hypersetup{%
    hypertexnames=false,
    colorlinks=true,
    allcolors=blue,
    unicode=true
}

\newcommand{\phantomsubfloat}[1]{%
    {%
        \captionsetup[subfloat]{farskip=0pt,captionskip=0pt}
        \captionsetup[subfigure]{labelformat=empty}
        \subfloat{#1}
    }
}

\crefname{section}{Sec.}{Sec.}
\crefname{appendix}{App.}{App.}

\begin{document}

\title{Universal quantum computation via scalable measurement-free error correction}

\author{Stefano Veroni}
\affiliation{PlanQC GmbH, M\"{u}nchener Str. 34, 85748 Garching, Germany}

\author{Alexandru Paler}
\affiliation{PlanQC GmbH, M\"{u}nchener Str. 34, 85748 Garching, Germany}
\affiliation{Aalto University, 02150 Espoo, Finland}

\author{Giacomo Giudice}
\email{giacomo.giudice@planqc.eu}
\affiliation{PlanQC GmbH, M\"{u}nchener Str. 34, 85748 Garching, Germany}

\date{\today}

\begin{abstract}
    We show that universal quantum computation can be concretely made fault-tolerant without mid-circuit measurements.
    To this end, we introduce a measurement-free deformation protocol of the Bacon-Shor code to realize a logical $\mathit{CCZ}$ gate.
    Combined with a fold-transversal logical Hadamard gate, this enables a universal set of fault-tolerant operations using only transversal gates and qubit permutations.
    For the purpose of benchmarking under circuit-level noise, we develop an efficient method to simulate non-Clifford circuits with a small number of Hadamard gates.
    Separately, we demonstrate that certain CSS codes can be concatenated without measurements or having to rely on a universal logical gate set.
    This is made possible by means of a resource-efficient gadget---termed the `disposable Toffoli gadget'---that realizes the error-correcting feedback.
    Then, under concatenation of the Bacon-Shor code, we observe a fault-tolerance threshold at a circuit-level depolarizing noise rate of approximately $0.12\,\%$.
    Together, the deformation and concatenation protocols outline a blueprint for a fully fault-tolerant architecture without any feed-forward operation, particularly suited to state-of-the-art neutral-atom platforms.
\end{abstract}

\maketitle

\section{Introduction}
\label{sec:introduction}

Given the inherently noisy nature of quantum hardware, it is crucial to develop approaches that suppress or limit the impact of errors.
One possible pathway is quantum error correction (QEC), which provides a means to detect and correct errors during the execution of quantum algorithms~\cite{terhal2015, campbell2017}.
At the heart of QEC lies the encoding of quantum information into a specific subspace within the broader physical Hilbert space---often achieved through a non-local encoding---allowing the system to be resilient against local errors. 
This redundancy, while effective, also significantly increases the resource overhead required for implementing QEC, both in terms of time and space. 
As a result, the choice of the error-correction protocol should be tailored to the specific quantum computing hardware, both to minimize overhead and to best exploit the platform's characteristics.
Despite these complexities, recent experimental breakthroughs have marked remarkable progress in QEC across different architectures, such as superconducting qubits~\cite{google2021, krinner2022, zhao2022, google2023, gupta2024, caune2024-riverlane, google2024, eickbusch2024, lacroix2024}, trapped ions~\cite{egan2021, ryan-anderson2021, hilder2022, postler2022, wang2023, postler2023, pogorelov2024, ryan-anderson2024-quantinuum, dasilva2024, berthusen2024_quantinuum}, nitrogen vacancies in diamonds~\cite{abobeih2022}, and neutral-atom arrays~\cite{bluvstein2024, reichardt2024, bedalov2024, rodriguez2024}.

In traditional QEC approaches, measurements of mutually-commuting correlators---known as \emph{error syndromes}---are used to inform a classical decoding algorithm, which then determines the corrective actions applied to the system, either through physical intervention or software-based tracking of errors~\cite{terhal2015}.
These measurements often correspond to \emph{stabilizer} operators, which commute with logical operators, such that the logical state remains unaffected~\cite{gottesman1997}. 

\begin{figure}[t!]
    \includegraphics[width=\linewidth]{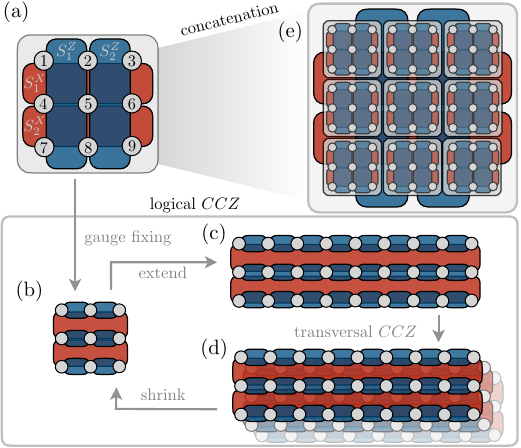}
    \phantomsubfloat{\label{fig:bacon-shor-stabilizers}}
    \phantomsubfloat{\label{fig:bacon-shor-gauged}}
    \phantomsubfloat{\label{fig:bacon-shor-extended}}
    \phantomsubfloat{\label{fig:bacon-shor-transversal}}
    \phantomsubfloat{\label{fig:bacon-shor-concatenation}}
    \vspace*{-2em}
    \caption{
        (a) The $3 \times 3$ Bacon Shor code, with red (blue) areas depicting the support of $X$-type ($Z$-type) stabilizer operators.
        (b) Any Bacon-Shor code has a gauge freedom. In particular, the pairwise $Z_i Z_j$ gauges along rows can be chosen to be $+1$, resulting in a Shor code.
        We call this the \emph{Shor} gauge.
        To perform a logical $\mathit{CCZ}$, each logical qubit must start from this gauge, and then be extended to a $3 \times 9$ Bacon-Shor code, as shown in (c).
        The logical $\mathit{CCZ}$ is then performed between logical qubits, by a transversal application of physical $\mathit{CCZ}$ gates, with permutations across the columns.
        (d) To return to the original $3 \times 3$ configuration, a shrink move is performed.
        (e) To increase the tolerance to errors, \emph{concatenation} of the same code is performed, where the data qubits of a code are composed of logical qubits at a lower level. 
    }
    \vspace*{-1em}
    \label{fig:bacon-shor-intro}
\end{figure} 

In some quantum computing platforms, such as those based on neutral atoms or trapped ions, the relatively slow nature of such measurements often hinders the efficiency of this syndrome extraction~\cite{pogorelov2024, postler2023, graham2023, singh2023, huie2023, lis2023, norcia2023, bluvstein2024}. 
An alternative approach, \emph{measurement-free} (MF) QEC, has emerged as a promising solution~\cite{paz-silva2010, li2012, crow2016, ercan2018, perlin2023, heussen2023, veroni2024}, replacing classical processing with a combination of unitary dynamics and a dissipative element needed to remove the entropy introduced by noise.
The latter is achieved through reset operations or a continuous supply of fresh ancilla qubits, which has recently been demonstrated in neutral-atom platforms~\cite{singh2022, norcia2024, gyger2024}.
In particular, the first fully \emph{fault-tolerant} (FT) proposal correcting a single error was achieved in Ref.~\cite{heussen2023}, by adapting Steane-type error correction techniques~\cite{steane1997}.
Later, some of the authors of this work showed that by using redundant syndrome information, alternative schemes can be formulated with a lower overhead in terms of both the number of qubits and gate count~\cite{veroni2024}.  

The notion of fault tolerance is essential for the design of an error-corrected quantum computer.
Broadly speaking, an FT implementation is capable of limiting error propagation, thus achieving a logical error rate that is lower than the physical error rate, as long as the latter remains below a certain threshold.
This is relevant both for the QEC round (which must not increase the noise level during its noisy execution), and for logical operations.
\emph{Transversal} operations are particularly appealing, as they are straightforwardly FT~\cite{terhal2015}.
Transversal operations correspond to a single layer of physical gates, each of which acts on at most one physical qubit of each logical qubit.
Therefore, errors can propagate between logical qubits, but remain correctable, since QEC is applied to each logical qubit block individually.

However, it is well known that no QEC code can have a transversal universal gate set~\cite{eastin2009-notransversal}, and, in particular, the only transversal gates for two-dimensional codes are unitaries belonging to the Clifford group~\cite{bravyi2013-konig}.
Several ways have been proposed to get around these no-go results.
These include magic state distillation~\cite{bravyi2005-magicstate, campbell2012-magicstate, campbell2017b-magicstate, gorman2017-magicstate, litinski2019-magicstate}, code switching between appropriate codes~\cite{anderson2014-codeswitch, beverland2021-codeswitch, butt2024-codeswitch}, using non-local connectivity between Bacon-Shor codes~\cite{yoder2017-baconshor-ccz}, implementing pieceable constructions where operations are intertwined with intermediate error correction~\cite{yoder2016}, or allowing for a loss of distance in concatenated codes~\cite{paetznick2013, jochym-oconnor2014, chamberland2017}.

The main theoretical challenges for MF QEC are two-fold: (\emph{i}) implementing universal quantum computation at the logical level, and (\emph{ii}) demonstrating scalability beyond small-distance codes.
We tackle these challenges in the following way.
We show (\emph{i}) that a \emph{universal gate set} is achievable with the Bacon-Shor code, by exploiting transversal operations and a MF code deformation procedure to realize a transversal $\mathit{CCZ}$ gate; and (\emph{ii}) that a universal gate set is \emph{not required} to perform MF QEC cycles at higher levels of concatenation, allowing for an efficient scaling up of the code distance.
During the completion of this work, the authors of Ref.~\cite{butt2024-mf} demonstrated MF code-switching between color codes to implement a universal set of logical gates.
Equipped with this universal set, it is then possible to concatenate the code with itself to reach higher distances.
In contrast, we show that the cost of concatenation can be lowered using resource-efficient gadgets, hence drastically increasing logical performance and scalability.

This work is organized as follows.
In \cref{sec:bacon-shor}, we review the Bacon-Shor code, which was shown to yield the best performance among currently devised MF schemes~\cite{veroni2024, heussen2023, perlin2023}.
In \cref{sec:logical-ccz}, we construct a MF procedure to \emph{deform} the code and exploit the protocol proposed in Ref.~\cite{yoder2017-baconshor-ccz} to implement a logical $\mathit{CCZ}$ gate, thus enabling MF universal quantum computation.
We then turn to address the second challenge in \cref{sec:concatenation}.
While this code deformation allows for a universal logical gates set, we take a different approach and implement MF QEC on concatenated codes \emph{without} the need for a universal gate set.
Inspired by Ref.~\cite{paz-silva2010}, we implement a feedback operation which we dub the \emph{disposable Toffoli gadget} and is significantly more efficient than using the aforementioned logical $\mathit{CCZ}$ construction.
The essence of this gadget is to \emph{unencode} ancilla qubits---storing stabilizer information---into a repetition code and then exploiting the (partial) transversality between repetition codes and a target code from the Calderbank-Steane-Shor (CSS)~\cite{calderbank1996,steane1996} family. 
Finally, in \cref{sec:outlook} we discuss potential advantages of this measurement-free approach, and how they are particularly suitable for neutral-atom platforms.

\section{The Bacon-Shor code}
\label{sec:bacon-shor}

In this section, we introduce the Bacon-Shor code (\cref{sec:bacon-shor-intro}), we summarize the construction for logical $\mathit{CCZ}$ used in Ref.~\cite{yoder2017-baconshor-ccz} (\cref{sec:logical-ccz-yoder}), and introduce some properties relating the Bacon-Shor code and the repetition code (\cref{sec:bacon-shor-repetition}), which we will use throughout this paper.

\subsection{Notation}
\label{sec:bacon-shor-intro}
The main properties of a QEC code are typically described by the triplet $\llbracket n, k, d \rrbracket$, where $n$ is the number of physical qubits, $k$ is the number of logical qubits, and $d$ is the code distance, i.e.\ the minimum number of single-qubit operations required to get from one codeword to another.
Additionally, we define $t$ as the number of \emph{tolerable} errors---i.e. all errors of weight smaller than or equal to $t$ can be corrected---satisfying $t \leq \lfloor \frac{d-1}{2}\rfloor$.

In this work, we will be focusing on the Bacon-Shor code~\cite{bacon2006,aliferis2007}.
This code essentially combines an $n_1$-qubit repetition code against $Z$ errors with an $n_2$-qubit repetition code against $X$ errors in an $n_1 \times n_2$ array of data qubits.
Labeling the qubits in this array with $(i, j)$, the stabilizers for the $\llbracket n_1 n_2, 1, \min(n_1, n_2) \rrbracket$ Bacon-Shor code are generated from
\begin{equation}
        S_i^X = \prod_{j=1}^{n_2} X_{i,j} X_{i+1,j}, \quad S_i^Z = \prod_{i=1}^{n_1} Z_{i,j} Z_{i,j+1}. \\
    \label{eq:bacon-shor-stabilizers}
\end{equation}
These are depicted in \cref{fig:bacon-shor-stabilizers} for the minimal error-correcting instance $\llbracket 9, 1, 3 \rrbracket$.
This specific instance is the main focus of this paper, as its MF QEC implementation, proposed in Ref.~\cite{veroni2024}, is particularly simple.

\begin{figure}
    \includegraphics[width=0.97\linewidth]{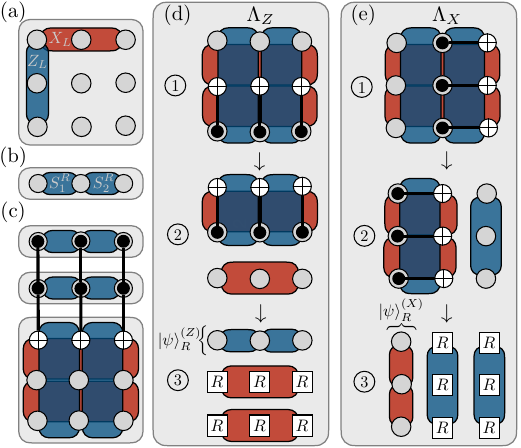}
    \phantomsubfloat{\label{fig:bacon-shor-logicals}}
    \phantomsubfloat{\label{fig:repetition}}
    \phantomsubfloat{\label{fig:bacon-shor-repetition-transversal}}
    \phantomsubfloat{\label{fig:bacon-shor-unencoding-z}}
    \phantomsubfloat{\label{fig:bacon-shor-unencoding-x}}
    \vspace*{-1.5em}
    \caption{
        (a) The logical operators of a Bacon-Shor code correspond to a single row ($X_L$) or column ($Z_L$).
        (b) A bit-flip repetition code is formed by pairwise $S^R_i = Z_{i} Z_{i + 1}$ stabilizers along a chain.
        (c) A transversal $\mathit{CCX}$ between a three-qubit repetition code and $d=3$ Bacon-Shor code.
        Note that this gate is only unidirectional.
        (d)--(e) The unencoding operation $\Lambda_Z$ ($\Lambda_X$) maps a Bacon-Shor code down to a bit-flip (phase-flip) repetition code.
        Intuitively, each column (row), supporting a separate representation of the relevant logical operator, is mapped to a qubit of the resulting repetition code.
        The stabilizers after each layer of gates are depicted.
    }
    \label{fig:unencoding}
\end{figure}

The logical operators are $X_L = \prod_j X_{i,j}$ and $Z_L = \prod_i Z_{i,j}$ for any value of $i$ or $j$ respectively, as shown in~\cref{fig:bacon-shor-logicals}.
As with all CSS codes, all logical Pauli operations are trivially transversal.
Additionally, the logical CNOT can be implemented transversally, by performing a CNOT between pairs of data qubits.
Additionally, the logical Hadamard gate is fold-transversal---it corresponds to a Hadamard gate on all data qubits and a relabeling of the physical qubits, which can be achieved with a reflection along the diagonal to swap $X$ and $Z$ stabilizers~\cite{aliferis2007}.

The Bacon-Shor code is a subsystem code~\cite{bacon2006, terhal2015}, such that multiple states encode the same logical codeword. 
These states are equivalent up to multiplication with \emph{gauge operators}, which commute with the stabilizers and logical operators, thus not affecting the logical information.
The gauge group is generated by the pairs $X_{i, j} X_{i+1, j}$ ($Z_{i, j} Z_{i, j+1}$) acting on qubits in the same column (row).
In what follows, we will be mainly interested in the \lq Shor gauge\rq\, of the code, where all $Z_{i, j} Z_{i, j+1}$ gauge operators have eigenvalue $+1$.
In this gauge, the codewords are equivalent to those of Shor's code~\cite{shor1995}.
In particular, in this gauge certain logical states become a tensor product of cat states 
\begin{equation}
    \label{eq:shor-codeword}
    \ket{\pm}_L^{(Z)} =
    \frac{1}{2\sqrt{2}}
    \left(\ket{000} \pm \ket{111} \right)^{\otimes 3}. 
\end{equation}
These states are particularly important, since they can be prepared fault-tolerantly without additional overhead~\cite{egan2021}.
We will denote the opposite gauge, in which all $X_{i, j} X_{i+1, j}$ have eigenvalue $+1$, as the `anti-Shor' gauge.
In this gauge, the states $\ket{0}_L$ and $\ket{1}_L$ have a similar structure as \cref{eq:shor-codeword}, this time as a tensor product of cat states along the columns instead of rows.
A mixed gauge can be chosen as well, as long as the gauge operators commute.
In particular, the rotated surface is a specific gauge choice of the Bacon-Shor code \cite{li2019-compass}.

\begin{figure}[t]
    \includegraphics[width=\linewidth]{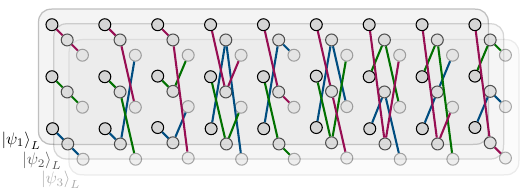}
    \vspace{-2em}
    \caption{
        Graphical depiction of the connectivity required for the transversal $\mathit{CCZ}_L \ket{\psi_1 \psi_2 \psi_3}_L$ between three logical states  $\ket{\psi_k}_L$, $k=1,2,3$, each encoded in the $3\times 9$ the Bacon-Shor codes, in the Shor gauge.
        Different colors correspond to different $\mathit{CCZ}$ gates between rows, and highlight that the permutations occur only across columns.
    }
    \label{fig:transversal-ccz}
\end{figure}

\subsection{Logical \texorpdfstring{$\mathit{CCZ}$}{CCZ}}
\label{sec:logical-ccz-yoder}
It was shown in Ref.~\cite{yoder2017-baconshor-ccz} that a $m \times m^k$ Bacon-Shor code in the Shor gauge supports a transversal logical $C^k Z_L$, up to a qubit permutation.
In particular, as shown in \cref{fig:transversal-ccz}, the logical $\mathit{CCZ}_L$ between three $3 \times 9$ logical qubits is implemented by a \emph{single layer} of 27 physical $\mathit{CCZ}$ gates 
\begin{equation}
    \label{eq:transversal-ccz}
    \mathit{CCZ}_L = \prod_{j=0}^8 \prod_{i=0}^2 \mathit{CCZ}_{
        (i,j),\,
        (i \oplus \lfloor j/3 \rfloor, j),\,
        (i\oplus j, \, j)
    },
\end{equation}
where it is assumed that $(i,j)$ indices start from $0$ for ease of notation, and $\oplus$ represents addition modulo $3$.

Since the Hadamard and $\mathit{CCZ}$ gates together define a universal gate set, it is then sufficient to demonstrate that these two gates can be performed fault-tolerantly.
As our starting point, we choose the $3 \times 3$ Bacon-Shor code, since it has a MF QEC implementation~\cite{veroni2024}, as well as a straightforward fold-transversal Hadamard gate.
However, certain technicalities need to be addressed.
The transversal realization of the logical Hadamard gate $H_L$ is only supported on a $m \times m$ patch of physical qubits, while the $\mathit{CCZ}_L$ gate in \cref{eq:transversal-ccz} requires a $m \times m^2$ patch.
Furthermore, the latter requires to be in the Shor gauge, which is not preserved under the application of $H_L$.
To address these issues, in \cref{sec:logical-ccz} we will demonstrate a fully MF protocol to implement the logical gate set: first, the code is mapped to the Shor gauge, by means of a \emph{gauge fixing} procedure (\cref{fig:bacon-shor-gauged}), then extended to a $3 \times 9$ code (\cref{fig:bacon-shor-extended}); now the logical $\mathit{CCZ}$ can be performed using \cref{eq:transversal-ccz}, cf.\ \cref{fig:bacon-shor-transversal}.
To return to the original code, we then shrink it down to a $3 \times 3$ code (\cref{fig:bacon-shor-gauged}).

\subsection{Relationship with the repetition code}
\label{sec:bacon-shor-repetition}

Before continuing with the logical protocol, we wish to outline some properties connecting CSS codes with repetition codes, which we will exploit later on.
The bit-flip (phase-flip) repetition code is defined on $n$ qubits by pairwise stabilizers $S^R_i = Z_{i} Z_{i + 1}$ ($S^R_i = X_{i} X_{i + 1}$), cf.\ \cref{fig:repetition}, and protects against $X$-type ($Z$-type) errors.
The respective logical codewords are 
$\ket{0/1}_R^{(Z)} = \ket{0/1}^{\otimes n}$ and $\ket{\pm}_R^{(X)} = \ket{\pm}^{\otimes n}$.
Generalizing the constructions proposed in Ref.~\cite{paz-silva2010}, throughout this work we exploit the fact that \emph{unidirectional} transversal gates exist between repetition codes and CSS codes.
In particular, a logical $\mathit{C^k X}$ ($C^k Z$) gate on a distance-$d$ code can be controlled from $k$ length-$d$ bit-flip repetition codes, as illustrated in \cref{fig:bacon-shor-repetition-transversal} for the Toffoli gate.
This gate is transversal, as it is composed of $d$ physical $\mathit{C^k X}$ ($C^k Z$) gates targeting $d$ qubits supporting the $X_L$ ($Z_L$) operator of the code.
As explained in \cref{sec:bacon-shor-repetition-extra}, the stabilizer structure is preserved.

Additionally, we introduce a gadget to convert a logical codeword to a codeword of the repetition code.
We call this an \emph{unencoding} gadget, and can be used when partial protection against either bit flips or phase flips is sufficient.
For an arbitrary logical state $\ket{\psi}_L = \alpha \ket{0}_L + \beta \ket{1}_L$, $|\alpha|^2 + |\beta|^2 = 1$, we introduce the unencoding operations
\begin{align}
    \Lambda_{X/Z} \,\ket{\psi}_L = \ket{\psi}_R^{(X/Z)} \otimes\ket{0}^{\otimes 6},
\end{align}
which converts it to $\ket{\psi}_R^{(X/Z)} = \alpha \ket{0}_R^{(X/Z)} + \beta \ket{1}_R^{(X/Z)}$.
Notice that these gadgets are not unitary, since information about the gauge choice is removed by the reset operations $R$.
The implementation of the unencoding gadgets for the $3 \times 3$ Bacon-Shor code is illustrated in \cref{fig:bacon-shor-unencoding-z,fig:bacon-shor-unencoding-x}, and is similar to the reverse of the encoding circuit. 
This is somewhat reminiscent of code morphing~\cite{vasmer2022}, albeit differing both conceptually and implementation-wise.
Effectively, the unencoding gadget $\Lambda_Z$ ($\Lambda_X$) maps the value of each column's $Z_L$ (row's $X_L$) on a qubit of the resulting repetition code---thus it is applicable to any gauge choice of the Bacon-Shor code, e.g. the rotated surface code. 
For generic CSS codes, such gadgets can be implemented, albeit with a larger overhead, by repeatedly performing Hadamard tests on different logical strings of the logical operators, and performing a majority vote, as was proposed in Ref.~\cite{paz-silva2010}.

By combining the unencoding gadget with the property in \cref{fig:bacon-shor-repetition-transversal}, we can engineer a FT feedback operation without measurements: a logical operation can be conditioned on another logical state, at the price of having to unencode the latter to a repetition code.
This can still be useful in many situations, such as when the control state is later discarded.
For instance, this will be used later on to perform the gauge fixing procedure detailed in \cref{sec:logical-ccz}, and in \cref{sec:concatenation} to construct an efficient feedback operation for concatenated codes.

\section{Measurement-free fault-tolerant logical \texorpdfstring{$\mathit{CCZ}$}{CCZ}}
\label{sec:logical-ccz}

\begin{figure*}[t]
    \includegraphics[width=\linewidth]{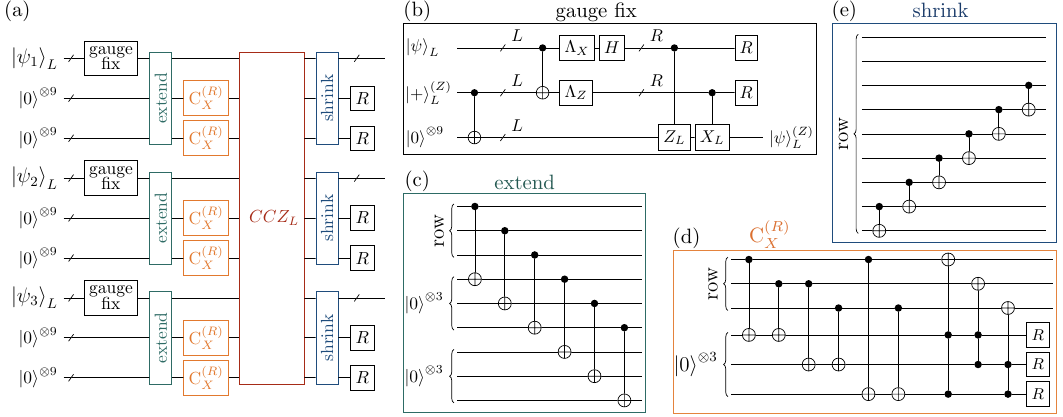}
    \phantomsubfloat{\label{fig:ccz-full}}
    \phantomsubfloat{\label{fig:ccz-results-gauge-fixing-teleportation}}
    \phantomsubfloat{\label{fig:ccz-results-gauge-fixing-extend}}
    \phantomsubfloat{\label{fig:repetition-correction}}
    \phantomsubfloat{\label{fig:ccz-results-gauge-fixing-shrink}}
    \vspace*{-1em}
    \caption{
        (a) Components of the protocol for the $\mathit{CCZ}_L$ gate between $3 \times 3$ Bacon-Shor codes.
        (b) Measurent-free gauge fixing by teleportation.
        A logical Bell state $(\ket{00}_L + \ket{11}_L) / \sqrt{2}$ is first prepared in the Shor gauge.
        Note that, because the $\ket{+}_L$ is prepared in the Shor gauge, the last logical qubit does not need to be prepared in $\ket{0}_L$.
        This auxiliary state is then entangled with the logical state, and then MF teleportation using $\mathit{CX}$ and $\mathit{CZ}$ gates is performed, by first unencoding to the repetition code, cf.~\cref{sec:bacon-shor-repetition}.
        (c) The extend gadget deforms a $3 \times 3$ Bacon-Shor code to its $3 \times 9$ variant, by exploiting the Shor gauge in the previous step.
        (d) A measurement-free error-correction round for the repetition code, used to correct for single bit flips on each triplet of each row of the logical states.
        (e) The extend gadget is reversed by a shrink gadget, bringing the logical state back to a $3 \times 3$ Bacon-Shor code.
    }
    \label{fig:logical-ccz}
\end{figure*}

In this section, we discuss in detail the protocol for a fault-tolerant $\mathit{CCZ}_L$, schematically illustrated in \cref{fig:bacon-shor-intro}.
For the distance-$3$ Bacon-Shor code, this corresponds to the protocol in \cref{fig:logical-ccz}.
We describe in more detail this protocol in \cref{sec:ccz-design} and benchmark it under depolarizing noise in \cref{sec:logical-performance}.

\subsection{Circuit design}
\label{sec:ccz-design}

To apply \cref{eq:transversal-ccz}, the Shor gauge must be first enforced on the code.
In general, this is necessary in a FT quantum computation even if we initialized logical states in the correct gauge, since the $H$ gate exchanges the Shor and anti-Shor gauge.
Furthermore, the QEC circuit proposed in Ref.~\cite{veroni2024}, and later adapted in \cref{sec:concatenation}, corrects every single-qubit error up to a gauge operator.
From a measurement-free perspective, fixing the gauge is somewhat more challenging than correcting single errors, as we must account for both gauge flips as well as possible single-qubit errors, so the heuristics developed in Ref.~\cite{veroni2024} are not applicable.
To gauge a logical state, we propose two different alternatives: (\emph{i}) a \emph{teleportation} protocol and (\emph{ii}) \emph{Steane-type gauge fixing}.
Both have similar performances under depolarizing noise, so we focus on the former, while the latter is relegated to \cref{sec:gauge-fixing-extra}.

In the teleportation protocol, the logical state is moved to a fresh register, which is initialized in the correct gauge.
The gauge information of the original state is removed by the unencoding gadget.
Its MF variant is illustrated in \cref{fig:ccz-results-gauge-fixing-teleportation}, and requires two additional logical registers.
Note that it can also correct single bit-flip errors, and has the added benefit of converting potential leakage errors into computational errors.

To extend a Bacon-Shor code in the Shor gauge from $3 \times 3$ to $3 \times 9$, it is sufficient to perform two consecutive transversal CNOTs from the code to a $\ket{0}^{\otimes 9}$, cf.~\cref{fig:ccz-results-gauge-fixing-extend}.
However, this may propagate bit flips that may lead to uncorrectable errors, so to make the procedure FT we introduce bit-flip corrections $\mathrm{C}_X^{(R)}$ along the rows, cf.~\cref{fig:repetition-correction}.
To shrink back a logical state to the $3 \times 3$ code, it is then sufficient to perform a series of CNOTs, illustrated in \cref{fig:ccz-results-gauge-fixing-shrink}.
Since the circuits act independently on each row, phase-flips cannot propagate to different rows and remain correctable.
In principle, such schemes can be generalized to any $3 \times 3^k$ Bacon-Shor code, to directly perform transversal $C^k Z$ gates~\cite{yoder2017-baconshor-ccz}.

\subsection{Numerical results}
\label{sec:logical-performance}

We benchmark the scheme in \cref{fig:logical-ccz} against the hardware-agnostic depolarizing noise, where after each gate, a random Pauli error is applied with probability $p$.
That is, given the set of Pauli strings of length $\ell$, i.e. $\mathcal{P}_\ell = \{I,X,Y,Z\}^{\otimes \ell}$, any $\ell$-qubit gate is followed by the channel 
\begin{equation}
    \mathcal{E}(\rho) = (1 - p) \rho + \frac{p}{|\mathcal{P_\ell}|-1} \sum_{\mathclap{\substack{P \in \mathcal{P}_\ell \\ P \neq I^{\otimes \ell}}}}{P \rho P^\dagger}.
    \label{eq:depolarizing}
\end{equation}
Considering that the circuits contain more than $80$ qubits and 27 physical $\mathit{CCZ}$ gates, both state vector simulators as well as Clifford simulators are not suitable.
Nevertheless, the logical states contain very few non-zero amplitudes in the computational basis, so the state vector is extremely \emph{sparse}.
In particular, the worst-case scenario is $\ket{\pm}_L$ in the anti-Shor gauge, since it is the eigenstate of $6$ independent $XX$ gauge operators and $X_L$, leading to $2^{6+1} = 128$ non-zero amplitudes.
If one simply keeps track of these non-zero entries---much as one would do with ``pen and paper'' calculation---the simulations can be performed with very moderate computational resources.

For this task, we develop \texttt{SparseStates.jl}~\cite{sparsestates} a simulation package in Julia which can efficiently simulate circuits with few \emph{branching} gates, i.e. gates that create superpositions in the computational basis (which are only $H$ gates in our case).
This allows us to compute thousands of circuit samples per second with circuits involving more than $100$ qubits and hundreds of Toffoli-like gates.
To the best of our knowledge, only Ref.~\cite{jaques2022} implements a similar method, based on hash maps, and with further gate rearrangements for the optimized simulation of quantum algorithms.
Instead, our state-vector implementation is array-based, which allows us to perform most operations in place---avoid unnecessary memory allocations---and opening up the possibility to exploit vectorized instructions available on most general-purpose processors.

To assess the performance of a protocol ideally realizing an operation $U$ at the logical level, we compute the average logical fidelity~\cite{pedersen2007}, by averaging over all different combinations of Pauli eigenstates $\mathcal{S}_\ell = \{\ket{\pm X}, \ket{\pm Y}, \ket{\pm Z}\}^{\otimes \ell}$ as input states~\footnote{%
    To reduce the computational cost, for numerical simulations of $\mathit{CCZ}_L$ we exploit the permutational invariance and the local $Z_L$ symmetry, and adjust the weights in \cref{eq:fidelity-ccz} accordingly.
}. 
To account for the degeneracy of the gauge choice, we expand the overlap between each target state and the resulting state $\rho_\sigma$ as a sum over expectation values of logical Pauli strings
\begin{equation}
    F_{U} = \frac{1}{|\mathcal{S}_\ell|} \! \sum_{\ket{\sigma} \in \mathcal{S}_\ell} \! \braket{\sigma | U^\dagger \rho_\sigma U | \sigma}_L
    = \sum_{\mathclap{\substack{\ket{\sigma} \in \mathcal{S}_\ell \\ P \in \mathcal{P}_\ell}}} c_\sigma^{(P)} \, \mathrm{Tr} \left(P_L \rho_{\sigma} \right),
    \label{eq:fidelity-ccz}
\end{equation}
where $\ell$ is the number of logical qubits $U$ acts on and $c_\sigma^{(P)} = \tfrac{1}{|\mathcal{S}_\ell|} \braket{ \sigma | U^\dagger P U | \sigma}_L$.
Different Kraus operators are sampled according to \cref{eq:depolarizing}, simulating up to $2^{24}$ realizations or $2^{10}$ failures per state, whichever comes first.

\begin{figure}
    \includegraphics[width=\linewidth]{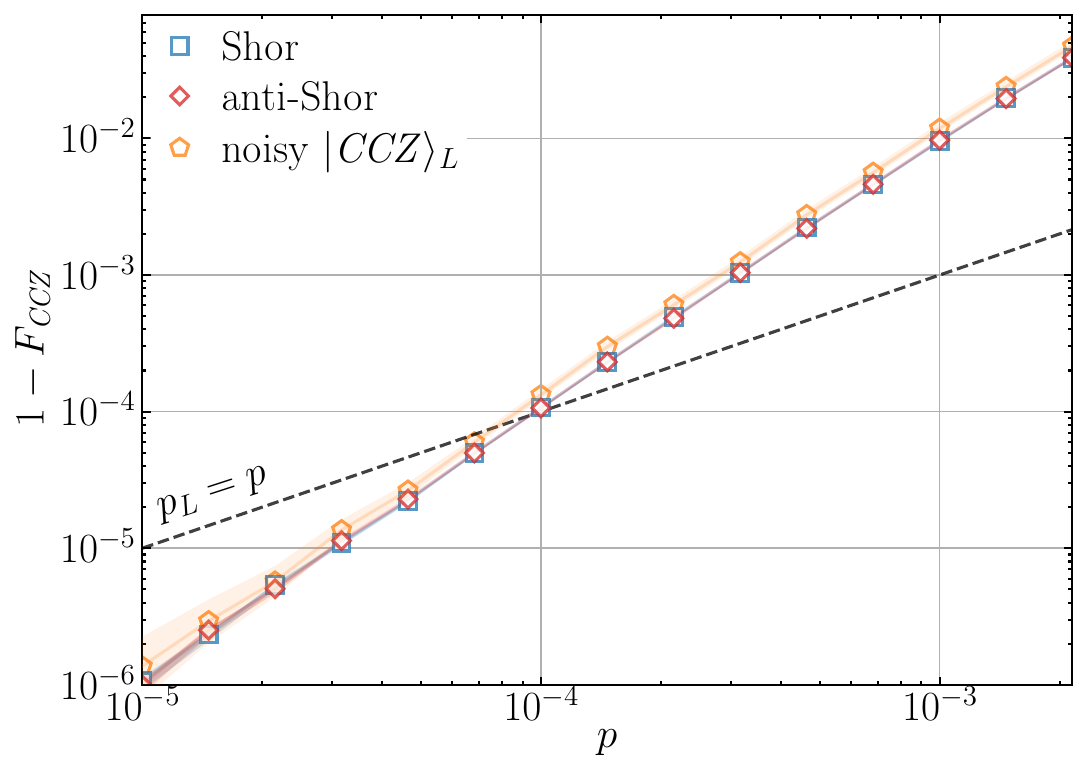}
    \vspace*{-2em}
    \caption{
        Average error rate of the logical $\mathit{CCZ}$ gate under depolarizing noise, computed using \cref{eq:fidelity-ccz}.
        The performance is benchmarked for initial states encoded without errors in the Shor (blue) or anti-Shor (red) gauges.
        Starting from $\ket{+++}_L$, we can prepare the non-Clifford resource state $\ket{CCZ}_L$.
        We therefore show the fidelity (orange) of its preparation using a noisy FT preparation in the Shor gauge, and without any gauge-fixing.
        Shaded regions correspond to $\qty{99}{\percent}$ confidence intervals.
    }
    \vspace*{-1em}
    \label{fig:ccz-results}
\end{figure}

The numerical results for the $\mathit{CCZ}_L$ gate are presented in \cref{fig:ccz-results}, for two different gauge choices.
However, these choices have minimal impact on the results, owing to the gauge-fixing procedure (see also~\cref{sec:gauge-fixing-extra}).
The quadratic scaling law of the logical error confirms the fault-tolerance of the scheme. 
Additionally, we show the performance of preparing the $\ket{\mathit{CCZ}}_L = CCZ_L\ket{+++}_L$ magic state, with a noisy FT initialization the Shor gauge, and hence not requiring any gauge fixing.
This state can serve as a non-Clifford resource, and may be useful for alternative constructions, for instance as the starting point of a distillation protocol to obtain better logical fidelities.
Interestingly, the preparation of $\ket{\mathit{CCZ}}_L$ exhibits a modest performance penalty, despite the absence of gauge fixing.
This can be attributed to the noisy preparation and the inherent asymmetry in the $\mathit{CCZ}_L$ protocol.
States like $\ket{+++}_L$ are maximally vulnerable to phase errors, and most subroutines in the protocol let single phase flips propagate, as they do not lead to uncorrectable errors.
However, at higher orders in $p$, multiple phase flips can accumulate, leading to logical phase errors.

Finally, we note that a MF code switching protocol was recently proposed by Ref.~\cite{butt2024-mf}, to realize a logical $T$ gate and achieving a breakeven fidelity of $\num{2.6e-4}$.
This figure is, broadly speaking, comparable to the numerical results presented here.
Notably, decomposing a $\mathit{CCZ}$ gate requires $6$ $CX$ gates and $7$ $T$ gates~\cite{nielsen-chuang, shende2008-cnots}.
The relative utility of these non-Clifford gates depends on the algorithmic context.

\section{Measurement-free concatenation}
\label{sec:concatenation}

Having demonstrated all the ingredients for FT and MF universal computation, in this section we show that the Bacon-Shor code can be scaled up by means of concatenation, thus achieving exponential suppression of noise.
In subsection~\ref{sec:concatenation-general} we introduce code concatenation and how it applies to MF QEC.
Then, in~\cref{sec:concatenation-design} we present an optimized QEC protocol for the concatenated Bacon-Shor code.
We do not exploit the deformation procedure developed in \cref{sec:logical-ccz}, but rather introduce a gadget to perform MF feedback operations.
We call this the \textit{disposable} Toffoli gadget, since it effectively acts as a transversal gate as long as the control qubits do not need to be protected or preserved, such as the case of ancilla qubits.
Numerical benchmark are presented in~\cref{sec:concatenation-performance}.

\subsection{Code concatenation}
\label{sec:concatenation-general}

We explore repeated \emph{concatenation} of the same code, which we denote as 
\begin{equation}
    \mathcal{C}_N = \underbrace{\mathcal{C} \circ \dots \circ \mathcal{C}}_{N~\text{times}},
\end{equation}
such that the first layer of concatenation is $\mathcal{C}_1 \equiv \mathcal{C}$.
The concatenation operation $\circ$ is a way of combining two codes: the logical qubits of the \emph{inner} code are used as physical qubits of the \emph{outer} one, cf.~\cref{fig:bacon-shor-concatenation}.
By recursively applying this operation, we can then construct a $N$-layer concatenated code.

Concatenation is a modular way of increasing the code distance.
Indeed, by concatenating $N$ times a CSS code of distance $d$, the resulting distance is $d_N = d^{N}$~\cite{terhal2015}.
The number of errors $t_N$ that we can correct depends on the decoding strategy.
Since in MF QEC decoding and corrections are operated by noisy quantum circuits rather than via classical processing,
the complexity of the available feedback operations is in practice limited.
In the simplest decoding strategy each unit operates autonomously, decoding each inner code individually, \emph{layer by layer}.
This still achieves an exponential suppression, albeit slower, of errors at the cost of a polynomial increase in the size of the error-correction circuit.
In a FT protocol adopting this strategy, any $t$ faults at the lower level of concatenation are correctable, hence we have the recurrence relation $t_N = (t + 1) (t_{N - 1} + 1) - 1$, which yields 
\begin{equation}
    t_N = (t + 1)^{N} - 1. \label{eq:t_N}
\end{equation}
For a base code correcting a single error ($t = 1$), we obtain the sequence $t_N = 1, 3, 7, 15, \dots$ for the first layers of concatenation.
This corresponds to an \emph{effective distance} $d_N^\mathrm{eff} = 2t_N + 1$,  exponentially worse than the best case $d_N$, cf.~\cref{tab:properties}.
Nevertheless, one still has an exponential suppression of errors by increasing $t_N$, guaranteeing that a threshold exists~\cite{knill1998, aharonov1998, aliferis2005-exRec}.
To understand this, we consider a circuit-level error model, where errors occur at every gate independently with probability $p$.
Defining the logical failure rate at the $N^\mathrm{th}$ level of concatenation as $p_N$, with $p_0 \equiv p$, we have $p_{N+1} \sim C p_N^{t+1}$, where $C > 1$ is a constant, corresponding to the number of combinations of $t + 1$ faults leading to a logical error.
Therefore, if $p < p_\mathrm{th} = C^{-1/t}$, then recursively $p_{N+1}<p_{N} \; \forall N$, hence the suppression of the error rate is faster than the circuit growth and a threshold is attained.

To achieve the upper bound $t_N = \lfloor \frac{d^N-1}{2}\rfloor$, decoding strategies exist, but require message-passing between different layers~\cite{poulin2006}.
To see how layer-by-layer decoding is not optimal, consider the concatenation of a three-bit classical repetition code with itself.
This is equivalent to a nine-bit repetition code, which can correct up to four errors.
However, by decoding each inner code individually, we can correct only up to three errors, as two errors in two subcodes can lead to the wrong logical state.

One can see layer-by-layer MF correction as a \emph{decoding} trade-off: we trade fault-tolerance for lower \emph{latency}.
The latter includes: (1) \emph{measurement latency} -- the time to perform the quantum measurements; (2) \emph{decoding latency} -- the time to communicate the syndromes to the decoder, decode the syndromes, and then apply or track the corrections.
Feed-forward (FF) decoding uses measurements and classical computation to achieve the $t_N$ upper bound, but MF decoding can speed up the logical performance by allowing the decoding units to operate autonomously.
This addresses both measurement and decoding latencies, at the price of sub-optimal decoding and lower error tolerance.

Regarding the measurement latency, MF decoding is attractive for several platforms, including trapped ions and neutral atoms, as it completely side-steps the issue of having slow measurements, which is one of the main bottlenecks in a QEC round~\cite{pogorelov2024, postler2023, graham2023, singh2023, huie2023, lis2023, norcia2023, bluvstein2024}.
In \cref{sec:implementation} we analyze the threshold measurement time for which MF decoding achieves the same fault-tolerance as a standard FF protocol, in the case of neutral-atom arrays. 
We estimate that the measurement latency in current state-of-the-art hardware is a significant bottleneck, such that MF schemes could outpace their conventional counterpart for small to mid-distance codes ($d^\mathrm{eff} \lesssim 15$).

The MF protocol avoids the decoding latency as well.
The fastest decoders have been shown to operate in real-time only for small distances~\cite{google2024, caune2024-riverlane}.
Real-time decoding at larger distances with many logical qubits is extremely challenging, and an active field of research~\cite{battistel2023real, skoric2023parallel, liyanage2024fpga}.
It generally requires trading tolerance for speed, too: faster decoding, such as union-find~\cite{delfosse2021almost}, has lower thresholds.

\begin{table}
    \def\arraystretch{1.25}
    \setlength{\tabcolsep}{0.5em}
    \begin{tabular}{c|cccc}
        Code & $\llbracket n, k, d \rrbracket$ & $d^\mathrm{eff}$ & $t$ & pseudothreshold \\
        \hline
        $\mathcal{C}_1$ & $\llbracket 9, 1, 3 \rrbracket$ & $3$ & $1$ & $\num{3.84e-3}$ \\
        $\mathcal{C}_2$ & $\llbracket 81, 1, 9 \rrbracket$ & $7$ & $3$ & $\num{1.95e-3}$ \\
        $\mathcal{C}_3$ & $\llbracket 729, 1, 27 \rrbracket$ & $15$ & $7$ & $\num{1.50e-3}$ \\
    \end{tabular}
    \caption{
        Key properties of the first layers of concatenation, the effective distance $d^\mathrm{eff}$, the number of correctable faults $t$ ($d^\mathrm{eff} = 2 t + 1$) and pseudothresholds obtained from numerical simulations with depolarizing noise.
    }
    \label{tab:properties}
\end{table}

\subsection{Circuit design}
\label{sec:concatenation-design}

\begin{figure*}[t]
    \includegraphics[width=\linewidth]{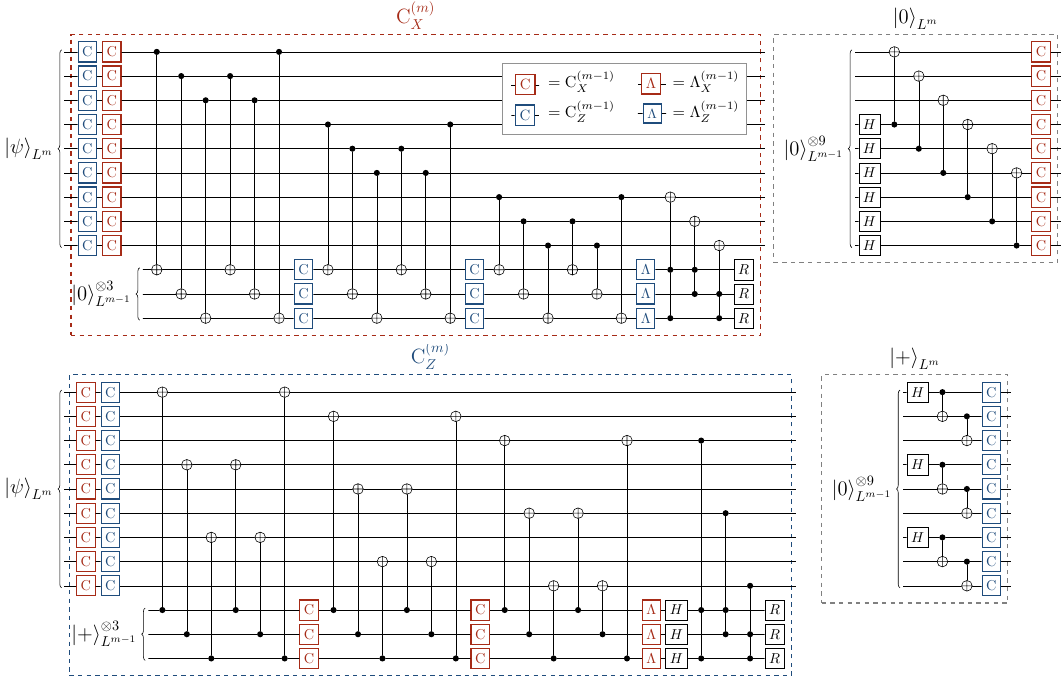}
    \vspace{-1.5em}
    \caption{
        Measurement-free fault-tolerant QEC implementation for the concatenated Bacon-Shor code, composed of a $\mathrm{C}_X^{(m)}$ and a $\mathrm{C}_Z^{(m)}$ subcircuit correcting for $X$ and $Z$ errors, respectively, where $m$ is the layer of concatenation.
        At the lowest layer $\mathcal{C}_1 \equiv \mathcal{C}$, intermediate error correcting subcircuits $\mathrm{C}^{0}_{X/Z}$ and unencoding operations $\Lambda^{0}_{X/Z}$ are replaced by identities.
        The ancilla registers at higher layers of concatenation are initialized fault-tolerantly using the circuits on the right, with $\ket{0/+}_{L^0} \equiv \ket{0/+}$.
    }
    \label{fig:concatenation-circuits}
\end{figure*}

The starting point of our concatenation procedure is the MF error-correcting circuit for the $3 \times 3$ Bacon-Shor code proposed in Ref.~\cite{veroni2024}.
For each subcircuit $\mathrm{C}_X^{(m)}$ ($\mathrm{C}_Z^{(m)}$) correcting bit-flips (phase-flips), $Z$-type ($X$-type) stabilizers are extracted to ancilla qubits, as shown in \cref{fig:concatenation-circuits}.
Additionally, a third stabilizer $S_3 = S_1 S_2$ is extracted, to be FT against circuit-level noise.
For a more detailed discussion on how to choose the set of redundant stabilizers and determine the circuit's design, we refer the reader to Ref.~\cite{veroni2024}.
When concatenating this code with itself, we promote both physical qubits as well as ancilla qubits to logical qubits of the lower layer.

For the correction, we construct a gadget to to perform the feedback efficiently, exploiting the fact that the phase information in the ancilla qubits is unimportant.
We unencode the ancilla qubits to a length-$d_N$ repetition code  using the unencoding gadgets repeatedly (see~\cref{fig:bacon-shor-unencoding-z,fig:bacon-shor-unencoding-x}), to exploit the transversality of multi-controlled gates described in \cref{sec:bacon-shor-repetition} and shown in~\cref{fig:bacon-shor-repetition-transversal}.
We note that the involved physical operations either have disjoint supports or act on the same column (row).
As can be seen in \cref{fig:bacon-shor-logicals} these columns (rows) are orthogonal to the $X_L$ ($Z_L$) operator, thus the unencoding protocol $\Lambda_Z$ ($\Lambda_X$) is FT, as each error location can affect at most one qubit of the repetition code.
By partially unencoding the logical ancilla qubits to a repetition code, we can then perform transversal multi-controlled operations on the logical qubits.
This is well-suited to the MF context, since it reduces the overhead for MF feedback operations while preserving fault-tolerance---as phase errors on the ancillas are not transferred to the data qubits, and single bit flips result in at most a correctable bit flip on the data qubits.

Finally, when concatenating a desired code, we need to include the error-correcting rounds for the underlying layers, which we call \emph{subrounds}.
To ensure fault-tolerance, subrounds can be introduced after every single gate~\cite{aliferis2005-exRec}.
However, for the Bacon-Shor code we consider, we realize that in many locations this is not necessary.
In particular, we can allow errors on the ancilla qubits to propagate onto the logical qubit, as long as they correspond to a gauge of the code.
For instance, in $\mathcal{C}_X$ ($\mathcal{C}_Z$), a single phase (bit) flip after initialization of the first ancilla would result in an error $Z_{1,1} Z_{1,2}$ ($X_{1,1} X_{2,1})$ on the data qubits after the first two layers of CNOTs.
If we let these errors propagate further however, we would end up potentially with correlated errors along rows (columns) which are aligned with the logical $X_L$ ($Z_L$) operator, and would lead to an uncorrectable error.
Therefore, we just need to place a $\mathrm{C}_X$ ($\mathrm{C}_Z$) subround on the ancilla qubits after having extracted a $Z_{i,j} Z_{i+1,j}$ ($X_{i,j} X_{i,j+1}$) gauge.

\vspace*{-0.5em}
\subsection{Performance}
\label{sec:concatenation-performance}

We consider the performance of the circuits presented in~\cref{sec:concatenation-design} under depolarizing noise, defined in \cref{eq:depolarizing}.
We consider the first three layers of concatenation, summarized in \cref{tab:properties}.
The circuits are simulated with \texttt{stim}~\cite{stim}, which allows for large system-size simulations by using the stabilizer tableau formalism.

To convert Toffoli-like gates into a circuit composed solely of Clifford gates, we note that all ancillas are reset after use, such that we can separately sample their values as either $0$ or $1$.  
We therefore convert multi-qubit gates using the principle of deferred measurement
\begin{equation}
    \label{eq:delayed-measurement}
    \includegraphics[scale=0.75]{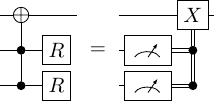},
\end{equation}
and including the corresponding noise channels defined by \cref{eq:depolarizing}, such that every Toffoli-like gate is followed by a three-qubit depolarizing noise channel.

Classical logic in \texttt{stim} is limited to exclusive-or, so, in order to benefit from the faster compiled samplers, we need to further delay the feedback operations.
To do so, we precompute a table associating each pair of virtual measurements to a list of future measurements that need to be flipped.
For example, for a quantum memory experiment with the $C_X^{(R)}$ gadget, the final part is replaced as follows  
\begin{equation}
    \label{eq:delayed-measurement-propagations}
    \includegraphics[scale=0.75]{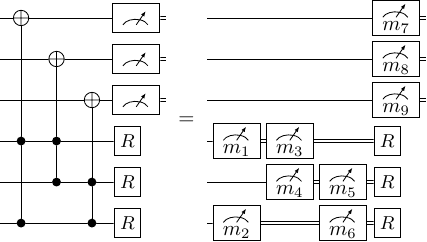},
\end{equation}
where $m_1 \wedge m_2$ flip $m_7$, $m_3 \wedge m_4$ flip $m_8$ and $m_5 \wedge m_6$ flip $m_9$.
In general, propagations of Pauli corrections are more complicated, but can still be computed efficiently, since the circuit is composed entirely of Clifford gates.
For this purpose, we use \texttt{stim}'s \texttt{FlipSimulator}.
This approach eliminates entirely all mid-circuit classical logic, enabling the use of the high-performance \texttt{CompiledMeasurementSampler}.
After sampling the feedback-free circuits, the measurement outcomes are sequentially updated using the precomputed table~\cite{zenodo}.
We then simulate noisy circuits stochastically for a variable number of cycles $N_\mathrm{C}$, for all eigenstates of the Pauli matrices, sampling up to $2^{24}$ realizations or $2^{10}$ failures---whichever occurs first.

\begin{figure}[t]
    \includegraphics[width=\linewidth]{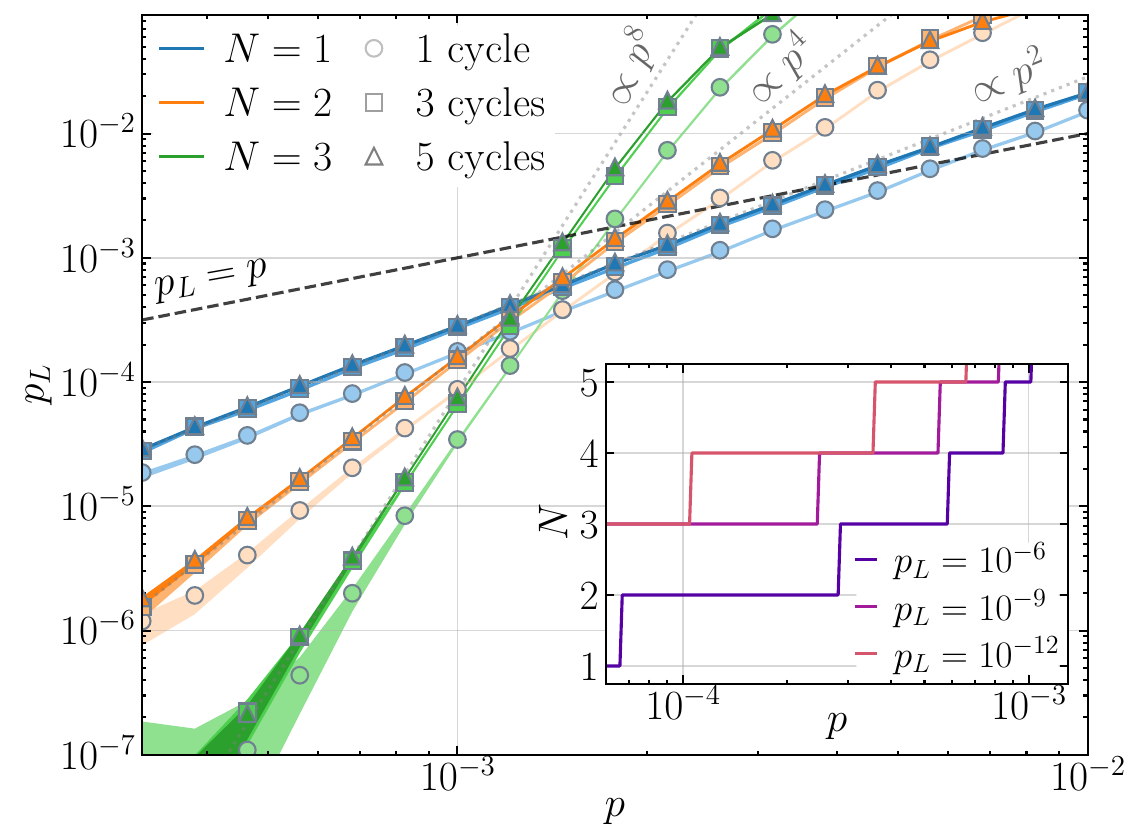}
    \vspace*{-2em}
    \caption{
        Logical error rate per cycle $p_L$ versus gate error rate $p$, for the first three layers of concatenation.
        The initial state is prepared without errors, and multiple QEC rounds are performed (markers).
        The solid lines correspond to the asymptotic estimates, extracted from a polynomial fit.
        From these estimates, we extrapolate the concatenation level $N$ required to reach different target logical error rates (inset).
        \vspace*{-0.2em}
    }
    \label{fig:concatenation}
\end{figure}

In \cref{fig:concatenation}, we show the logical error rate per cycle $p_L = 1 - F_I^{1/N_\mathrm{C}}$, calculated by averaging over all the eigenstates of the logical Pauli operators, cf.~\cref{eq:fidelity-ccz}.
The numerical results indicate that the failure rate of $\mathcal{C}_N$ scales asymptotically as $p^{t_N+1}$, as expected for a protocol that is robust against $t_N$ faults at the circuit level.
From this data, we extract an asymptotic threshold $p_\mathrm{th} \simeq \num{1.2e-3}$, and the pseudothresholds in \cref{tab:properties}, potentially relevant for real implementations.
Compared to universal concatenated constructions that rely on measurements~\cite{chamberland2017}, we note that our protocol has a comparable asymptotic threshold, despite having the overhead of being measurement-free.
For higher levels of concatenations, we can approximate the logical error rate sufficiently below threshold as $p_L(p) \sim  p_L(p_\mathrm{th}) \left( p / p_\mathrm{th} \right)^{t_N + 1}$.
As shown in the inset of \cref{fig:concatenation}, we can then estimate the level of concatenation required to reach a target $p_L$.

Overall, these results indicate that the MF error-correcting protocols can achieve below-breakeven logical error rates on near-term quantum devices, assuming physical error rates below $\qty{0.1}{\percent}$, generally considered within reach.
In particular, state-of-the-art trapped ion platforms have already demonstrated comparable error rates~\cite{moses2023,loschnauer2024,chen2024} and neutral atom arrays are making rapid progress towards this goal~\cite{ma2022,evered2023,finkelstein2024,tsai2024,muniz2025}.
Furthermore, a few layers of concatenation ($N \leq 3$) are sufficient to achieve a strong suppression of logical errors, within the regimes of interest for next-generation devices~\cite{preskill2025}.

\section{Outlook}
\label{sec:outlook}

We have constructed a fully fault-tolerant architecture for universal and scalable quantum computation that does not require any measurements.

First, by performing a measurement-free gauge fixing and code deformation, we have demonstrated a fault-tolerant protocol for the realization of a logical $\mathit{CCZ}$ gate in the Bacon-Shor code.
Together with the code's fault-tolerant initialization and transversal Hadamard gate, this enables measurement-free, fault-tolerant and universal operations.
These ideas of gauge fixing are very powerful, as well-known error-correction concepts such as lattice surgery and code deformation have been recast as a gauge-fixing procedure~\cite{vuillot2019}.
It is likely that in the future such concepts can be exploited in the measurement-free context, to devise more efficient universal logical operations.

\begin{figure}[t]
    \includegraphics[width=0.8\linewidth]{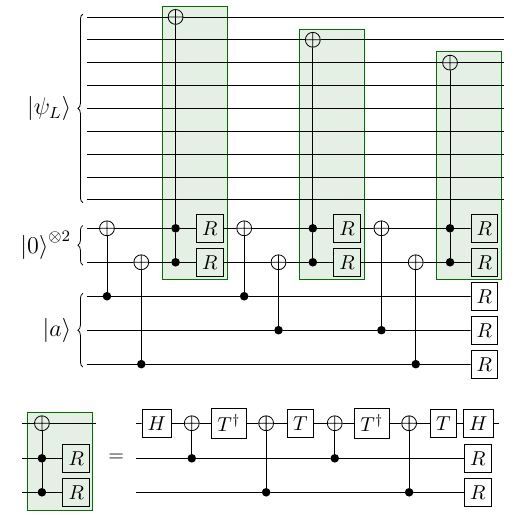}
    \vspace*{-1em}
    \phantomsubfloat{\label{fig:decomposition-a}}
    \phantomsubfloat{\label{fig:decomposition-b}}
    \caption{
        Decomposition of the correction step for the $3 \times 3$ Bacon-Shor code in~\cref{fig:concatenation-circuits} in terms of two-qubit gates.
        By copying the control qubits in the ancilla register $\ket{a}$, we can then use a Toffoli-reset gadget with fewer gates than the full decomposition, as already recognized in Ref.~\cite{heussen2023}.
    }
    \vspace*{1em}
    \label{fig:decomposition}
\end{figure}

Second, we propose to concatenate measurement-free implementations of small codes to realize higher-distance codes and thus achieve protection against an arbitrary number of faults.
Remarkably, these concatenated protocols do not require a universal gate set, as we use a gadget to perform the feedback operation in a hardware-efficient way. 
We stress that such constructions are not unique to the Bacon-Shor code, but could be adapted to other codes of the CSS family as well.
In particular, we note that the disposable Toffoli gadget is also compatible with the $d=3$ rotated surface code, without any modifications.
By performing numerical simulations, we find that the performance is competitive and can be considered for error-correcting experiments on near-term devices.
Furthermore, it can be expected that the protocols can be further tailored to a real hardware device.
For example, by exploiting biased noise, the performance of measurement-free protocols can be significantly enhanced~\cite{veroni2024}.
Furthermore, neutral atoms provide an ideal platform for the implementation of such protocols, as native $\mathit{CCZ}$ operations have been demonstrated~\cite{evered2023}.
In the case where such operations are not available, the protocols can be adapted without spoiling fault-tolerance.
As an example, as shown in \cref{fig:decomposition}, the correction operations can always be decomposed fault-tolerantly, by copying the control qubits to an auxiliary register, and performing a simplified decomposition of the feedback operation~\cite{heussen2023}.

Altogether, this architecture requires non-local, but heavily parallelizable operations.
This is an ideal setting for neutral-atom platforms, which have demonstrated parallel Clifford operations and parallel shuttling of multiple registers for the realization of transversal entangling operations~\cite{bluvstein2024, reichardt2024}.
Furthermore, continuous loading of atoms provides a reservoir of fresh qubits, which can emulate reset operations in a scalable way ~\cite{singh2022, norcia2024, gyger2024}.
Despite a potential performance overhead, measurement-free protocols can be a valid alternative to significantly speed up the \emph{logical clock rate}.
In \cref{sec:implementation}, we perform `back-of-the envelope' calculations that suggest that for fault-tolerant experiments in the foreseeable future ($d \lesssim 15$), measurement times would need to significantly decrease to match the clock times of their measurement-free counterpart.
This further motivates measurement-free error correction as a viable and scalable pathway towards fault-tolerant quantum computation. 

\begin{acknowledgments}
We acknowledge inspiring discussions with Gavin Brennen and Johannes Zeiher. AP acknowledges partial funding from the Defense Advanced Research Projects Agency [under the Quantum Benchmarking (QB) program under award no. HR00112230006 and HR001121S0026 contracts], and was supported by the QuantERA grant EQUIP through the Academy of Finland, decision number 352188. The views, opinions and/or findings expressed are those of the author(s) and should not be interpreted as representing the official views or policies of the Department of Defense or the U.S. Government.
\end{acknowledgments}

\appendix

\section{Transversality of controlled-gates between repetition codes and CSS codes}
\label{sec:bacon-shor-repetition-extra}

In this appendix, we show that a logical $\mathit{C^k X}$ ($C^k Z$) gate on a distance-$d$ CSS code can be controlled from $k$ length-$d$ bit-flip repetition codes, without affecting the stabilizer structure of the logical qubits involved.
These consist of $d$ physical $\mathit{C^k X}$ ($C^k Z$) gates targeting the $d$ qubits of the code supporting the $X_L$ ($Z_L$) operator.

\begin{figure*}
    \centering
    \includegraphics[width=\linewidth]{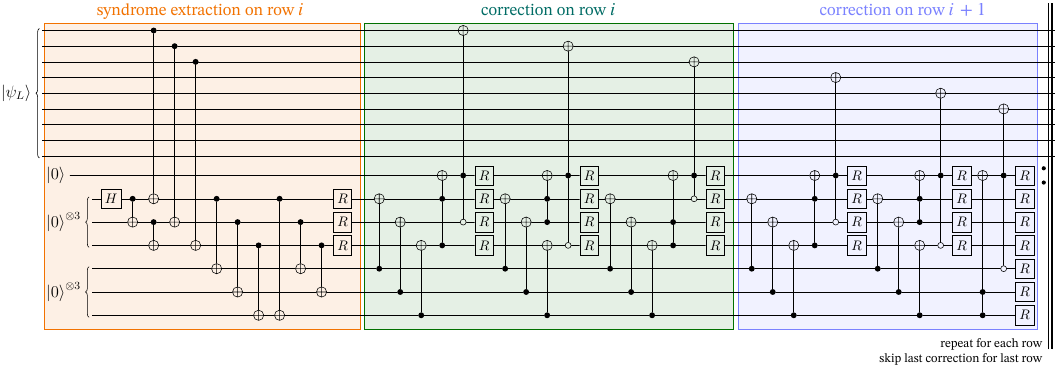}
    \vspace*{-2em}
    \caption{
        Gauge fixing by adapting Steane-type error correction.
        This circuit maps the $X$ errors of a single row into a three-qubit cat state, decodes it by a redundant extraction of the parities~\cite{veroni2024} (orange box), and, if a bit flip occurred, applies a $XX$ gauge operator between the current row (green box) and the next one (purple box).
        This circuit is repeated for all rows except for the last one, where remaining bit-flip errors are extracted and then corrected by applying only the first part of the corrections. 
    }
    \label{fig:gauge-fixing-steane}
\end{figure*}

For the $\mathit{CX}$ gate, using the well known propagation rules
\begin{equation}
    \begin{aligned}
        \includegraphics[scale=0.75]{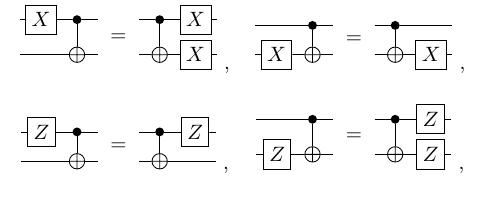}
    \end{aligned}
    \label{eq:propagations-cx}
\end{equation}
we realize that the only non-trivial propagation are the $S^Z$ stabilizers of the code overlapping with $X_L$, which get mapped to the stabilizers of the repetition code.
By definition, $X_L$ commutes with the stabilizers, meaning every overlap with $S^Z$ stabilizers involves an even number of qubits.
But an even product of distinct $Z$ operators can always be absorbed by the stabilizer generators of the repetition code.
In the case of a $\mathit{CZ}$ gate, the targets support the $Z_L$ operator, and equivalent results are obtained by using the identity $\mathit{CZ} = (I \otimes H) \mathit{CX} (I \otimes H)$.
By symmetry these arguments also hold for unidirectional gates on length $d$ phase-flip repetition codes controlled by distance-$d$ CSS codes.

We can now extend these result to three-qubit gates as well.
Again, using the propagation rules for the Toffoli gate
\begin{equation}
\label{eq:propagations-ccx}
    \begin{aligned}
        \includegraphics[scale=0.75]{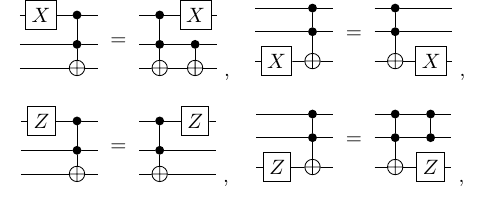}
    \end{aligned}
\end{equation}
the only non-trivial propagation arises from the $S^Z$ stabilizers of the code overlapping with $X_L$.
The even number of overlapping qubits results in an even number of $\mathit{CZ}$ operations between the controlling bit-flip codes, which reciprocally cancel. 
Similarly, in the $\mathit{CCZ}$ gate, the non-trivial propagation is due the intersection of $S^X$ stabilizers and $Z_L$, which again results in the propagation of an even number of $\mathit{CZ}$s, which reciprocally cancel out.

In general, the $C^k X$ ($C^k Z$) gates controlled on the bit-flip codes only have non-trivial propagation of the $S^Z$ ($S^X$) stabilizers of the code overlapping with $X_L$ ($Z_L$).
These lead to an even number of $C^{k-1} Z$ gates between the controlling bit-flip codes, which mutually cancel out, thus preserving the stabilizer structure.

\section{Steane-type MF gauge fixing}
\label{sec:gauge-fixing-extra}

As an alternative, we also propose a gauge fixing inspired by Steane-type MF QEC~\cite{steane1997, heussen2023}, which both corrects single-qubit bit-flips and enforces the Shor gauge, using fewer qubits if fast resets are available.

This protocol is illustrated in~\cref{fig:gauge-fixing-steane}.
The main idea is to treat each row one at a time, extracting each $ZZ$ gauge operator to detect bit-flips (orange box).
For each bit flip, we apply a $XX$ gauge operator between the current row (green box) and the next one (purple box).
This sets each of the $Z$-gauges to their $+1$ eigenvalue, and moves single bit-flip errors to the last row.
We can then perform a similar bit-flip correction on the last row to correct for a potentially remaining bit-flip.
As with Steane-type QEC, this scheme avoids the propagation of more than one error.

The scheme can be sped up, by extracting $ZZ$ gauge operators of the top and bottom rows simultaneously, and then applying the $XX$ gauge operator between said row and, respectively, the one below or above.
By repeating this process iteratively, between rows progressively closer to the middle, errors are moved to the middle row.
Finally, one round of extraction and correction on the middle row corrects a potentially remaining bit-flip.

\begin{figure}[h]
    \includegraphics[width=\linewidth]{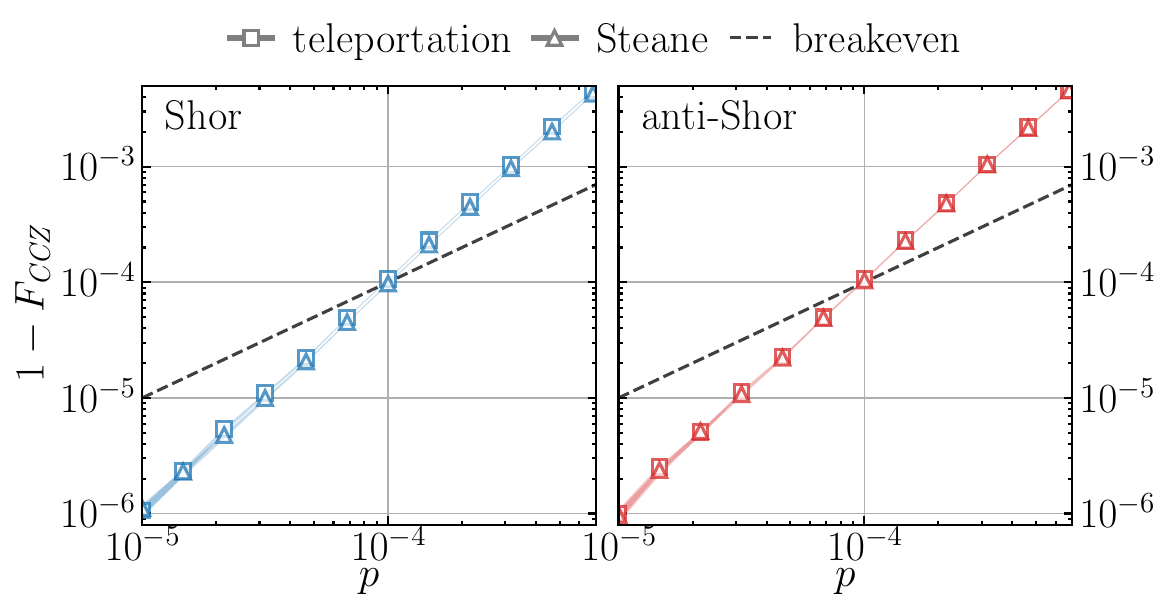}
    \vspace*{-2em}
    \caption{
        Logical error rate of the $CCZ_L$ gate, employing the teleportation or the Steane-type gauge-fixing protocols.
        This is computed using \cref{eq:fidelity-ccz}, for states encoded in either the Shor (left) or anti-Shor gauge (right).
    }
    \label{fig:ccz-results-gauge-fixing}
\end{figure}

A comparison of the two different gauge fixing protocols presented is shown in \cref{fig:ccz-results-gauge-fixing}, and exhibit very similar performance.
We note that the performance of the teleportation circuit is independent of the initial state's gauge, while the Steane-type gauge fixing has a marginally lower error rate when the gauge is correct.

\begin{figure*}[t]
    \includegraphics[width=\linewidth]{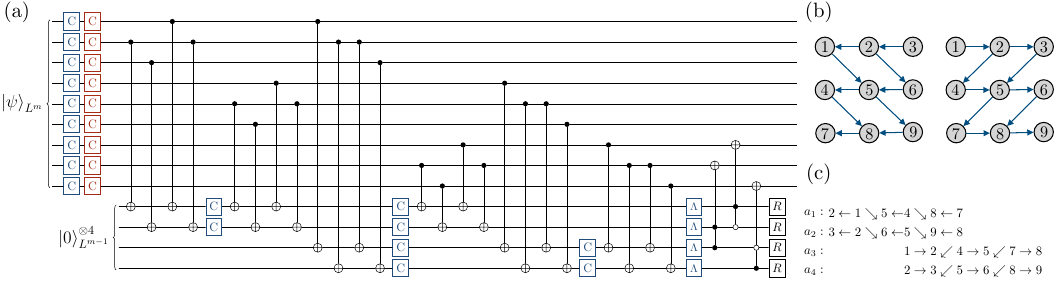}
    \phantomsubfloat{\label{fig:bacon-shor-variant-a}}
    \phantomsubfloat{\label{fig:bacon-shor-variant-b}}
    \phantomsubfloat{\label{fig:bacon-shor-variant-c}}
    \vspace{-1.5em}
    \caption{
        (a) Alternative measurement-free and fault-tolerant QEC implementation for the Bacon-Shor code, using four ancillae.
        The colored operations follow the same convention of~\cref{fig:concatenation-circuits}, indicating subcircuit error correction and unencoding routines.
        (b) Shuttling pattern for the first two (left) and last two (right) ancillae.
        (c) Shuttling schedule for the syndrome extraction, showing which data qubit each ancilla $a_i$ interacts with at a certain moment in time, and the direction of the shuttling in between.
        The order with which ancillae interact with data qubits is important to preserve fault-tolerance~\cite{veroni2024}. 
        The schedule proposed is chosen to maximize the number of parallel operations, while also avoiding the propagation of uncorrectable errors from the ancillae to the data qubits.
    }
    \label{fig:bacon-shor-variant}
\end{figure*}

\section{Practical considerations}
\label{sec:implementation}
In this section, we compare the implementation of the MF protocols and more conventional QEC protocols on a potential neutral-atom platform, from the point of view of logical clock time.
In particular, we assume a two-dimensional array of atoms, with parallel shuttling capabilities.
Since the concatenation procedure increases the footprint of the logical register, it is important to account for potentially long shuttling times.
For measurements, we consider the situation of either reconfigurable arrays with a separate readout zone~\cite{bluvstein2024}, or a static array with local imaging of ancillae, enabled for example by shelving of the data qubits to different atomic levels.
For simplicity, we only consider shuttling and measurement times, since these are typically much more important than the execution time of individual gates~\cite{bluvstein2022, bluvstein2024}.

With regards to shuttling, its speed is limited by the amount of heating it induces. 
Two main shuttling trajectories have been designed to address this problem, namely a constant-jerk (CJ) movement~\cite{bluvstein2022} or an acceleration profile based on shortcut-to-adiabaticity (STA) techniques~\cite{hwang2025}.
If an atom undergoes $K$ movements of distance $D_i$ each, the time taken by each movement is
\begin{equation}
    \Delta t_\mathrm{i} = \left({\alpha K D_i^2}\right)^f,
    \label{eq:shuttling-time}
\end{equation}
where $f=1/4\,(1/6)$ for CJ (STA) shuttling, and $\alpha$ is a constant that depends on the experimental setting and the trajectory. 
More details can be found at Refs.~\cite{bluvstein2022, hwang2025}.
On the other hand, with regards to mid-circuit measurements and feed-forward, current state-of-the-art experiments across various atomic species and qubit encodings have attained times ranging between $\qtyrange{1}{20}{\milli\second}$~\cite{bluvstein2024, lis2023, norcia2023, singh2023, huie2023, finkelstein2024, radnaev2024}.

For the MF QEC protocol, we assume the quasi-parallel extraction of individual stabilizers, since we can devise an alternative MF QEC protocol using four ancilla qubits instead of three, cf.~\cref{fig:bacon-shor-variant}.
The advantage of this protocol is that it avoids measuring the stabilizer $S_3$ (cf.~\cref{sec:concatenation-design}), which requires longer-range interactions, by measuring $S_1$ and $S_2$ twice.
Compared to \cref{fig:bacon-shor-concatenation}, it exhibits a slightly lower pseudothreshold, despite having more redundancy.
The total time required for a subcircuit $\mathrm{C}^{N}$ at $N$ levels of concatenation then can be broken down as 
\begin{equation}
    T_C^{(N)} = T^{(N)}_\mathrm{circ} + 5 T_C^{{(N-1)}},
    \label{eq:time-total}
\end{equation}
where $T^{(N)}_\mathrm{circ}$ is the time required for shuttling in between logical gates, whereas $5 T_C^{{(N-1)}}$ accounts for the error-correcting subcircuits $\mathrm{C}^{N-1}$.

In this scheme, each ancilla moves at most $K=6$ times: thrice horizontally and twice diagonally during the syndrome extraction (cf.~\cref{fig:bacon-shor-variant-b}), and at most once more horizontally for the application of the feedback Toffoli operation---the unencoding gadget can be realized with nearest-neighbor gates only.
Each horizontal movement, at the $N^\mathrm{th}$ concatenation level, covers a distance $D_i^{(N)}=\delta x \, d_{N-1}$, where $\delta x$ is the spacing between data qubits and $d_N = 3^N$;
diagonal movements are longer by a factor $\sqrt{2}$.
We note that the shuttling schedule presented is compatible with the constraints imposed by the acousto-optical-deflectors (AODs) used to steer atoms in neutral-atom hardware~\cite{schmid2024, constantinides2024}.
Then, by using \cref{eq:shuttling-time}, we have
\begin{equation}
    T^{(N)}_\mathrm{circ} = \sum_{i} \Delta t_i = \kappa \left(K\alpha 3^{2(N-1)} \delta x^2\right)^f,
\end{equation}
where $\kappa$ depends on the parallelization and relative size of the moves.
There are a total of 10 steps of parallel shuttling, covering syndrome extraction---which contributes eight steps as per cf.~\cref{fig:bacon-shor-variant-c}---and the feedback operation.
Six steps only include shorter horizontal moves, while in the remaining four the longest movement is diagonal.
Then, $\kappa = 6 + 4 \cdot 2^f$. 
Inserting this result in \cref{eq:time-total} we obtain
\begin{equation}
    T_\mathrm{C}^{(N)} = 5 T_\mathrm{C}^{(N-1)} + \kappa \left(K\alpha 3^{2(N-1)} \delta x^2\right)^f.
\end{equation}
Unraveling this recursive relation yields
\begin{align}
    T_\mathrm{C}^{(N)} &= \kappa \left(K\alpha \delta x^2\right)^f \sum_{n=0}^{N-1} 5^{n} 3^{2(N-1-n)f} \nonumber \\ &= 
    \kappa \left(K\alpha \delta x^2\right)^f\,
    \frac{5^N - 3^{2Nf}}{5-3^{2f}}.
\end{align}

For comparing with equivalent feed-forward (FF) implementations, we use a QEC code that is equivalent by the number of correctable errors $t_N$ rather than the distance, cf. \cref{eq:t_N}. 
Thus, we consider a logical qubit encoded in a $d_N^\mathrm{eff} \times d_N^\mathrm{eff}$ register, using some CSS code such as the rotated surface code, allowing for local stabilizer extractions without shuttling.
We assume that by correctly decoding the syndrome information, we can correct up to $t_N = \lfloor (d_N^\mathrm{eff}-1)/2\rfloor$ faults on the data qubits.
In this setting, a QEC cycle is then dominated by the measurement and feed-forward time $T_{FF}$.
In particular, for an extraction of a single type of syndromes, the measurements of the stabilizers is generally performed $d_N^\mathrm{eff}$ times to be robust against circuit-level noise.
Within this measurement time we also include the shuttling time to a separate readout zone, if needed, which can be non-negligible~\cite{bluvstein2024}.

The breakeven measurement and feed-forward time $T^\ast_{FF}$ for which the MF protocol with shuttling would match the FF implementation, for the same tolerance $t_N$, is then
\begin{equation}
    T^\ast_{FF} (N) = 2\kappa \left(K{\alpha \delta x^2}\right)^f
    \frac{5^N - 3^{2Nf}}{5-3^{2f}}
    \frac{1}{2^{N+1} - 1},
\end{equation}
where the factor 2 arises from the fact that the MF protocol measures $X$-type and $Z$-type stabilizers one after the other, whether we assume that they are measured in parallel in FF schemes.
With an estimate of $\delta x = \qty{4}{\micro\meter}$, and the experimentally demonstrated values~\cite{bluvstein2022, hwang2025}
\begin{align}
    \alpha_\mathrm{STA} &\approx 1.4\times 10^{-11}\,\mathrm{ms^6/\mu m^2}, \nonumber \\
    \alpha_\mathrm{CJ} &\approx 1.3\times 10^{-7}\,\mathrm{ms^4/\mu m^2},
\end{align}
we obtain 
\begin{equation}
    T^\ast_{FF} \simeq \qtylist{0.23; 0.64; 1.6}{\milli\second} 
\end{equation}
for $N = \numlist{1;2;3}$ with STA shuttling, and roughly twice as much with CJ shuttling.
Recalling that current state-of-the-art experiments have attained a $T_{FF}$ that ranges between $\qtyrange{1}{20}{\milli\second}$~\cite{bluvstein2024, lis2023, norcia2023, singh2023, huie2023, finkelstein2024, radnaev2024}, then the MF approach not only avoids the need for mid-circuit measurements but can shorten the QEC cycle time.

For the purposes of clarifying the impact of the experimental design on the performance of the protocol, we note that the QEC cycle time of the MF protocol is estimated to scale as
\begin{equation}
    T_C^{(N)} \propto \alpha^f \propto \frac{m^{\frac{1}{2}(1+f)} \,w_\mathrm{tr}^{1-f}}{U_\mathrm{tr}^{\frac{1}{2}(1-f)}},
\end{equation}
where $m$ is the mass of the used atom, $w_\mathrm{tr}$ and $U_\mathrm{tr}$ are the width and the depth of the trap, and recall that $f=1/4\,(1/6)$ for CJ (STA) shuttling.
This implies that by increasing the trapping depth and adopting lighter atomic species as ancilla qubits, a significant reduction in time could be achieved.

With regards to the comparison between MF and FF approaches, we note that, on the one hand, the FF implementation could be improved in principle by not applying $d$ rounds of syndrome extraction for every type of stabilizer, as recent research has shown that for a logical computation fewer rounds of measurements can be performed and fault-tolerance can still be preserved, at the price of a more complicated decoding~\cite{zhou2024}.
It remains to be seen if such approaches can be adapted to the measurement-free setting as well.
On the other hand, for a FF implementation of a logical computation it is important to take into account shuttling times to a separate readout zone, if required.
An increase in the number of logical qubits or in code distance necessarily increases the separation from the readout zone, which in turn lengthens the measurement time.
Considering that practically-relevant quantum algorithms will require hundreds to thousands logical qubits~\cite{dalzell2023}, future experiments will need to tackle such increasing overhead.
Moreover, the complexity of classical decoding increases with the distance of the QEC code, in turn providing additional overhead and latency.

\bibliography{library}

\end{document}